# Single Lateral Mode Mid-Infrared Laser Diode using Sub-Wavelength Modulation of the Facet Reflectivity

G. R. Nash*

College of Engineering, Mathematics and Physical Sciences, University of Exeter, Exeter, EX4 4QF, U.K.

J. L. Stokes, J. R. Pugh, and S. J. B. Przeslak

Department of Electrical and Electronic Engineering, University of Bristol, Bristol, BS8 1UB, U.K.

P. J. Heard

Interface Analysis Centre, University of Bristol, Bristol, BS8 1UB, U. K.

J. G. Rarity and M. J. Cryan

Department of Electrical and Electronic Engineering, University of Bristol, Bristol, BS8 1UB, U. K.

**ABSTRACT**

The characteristics of mid-infrared laser diodes have been investigated before and after the patterning of sub-wavelength metallic apertures on the emitting facet. Before modification of the facet the emitted spectrum consisted of a large number of peaks associated with different spatial modes, whereas afterwards the spectrum was dominated by a single peak. Simulations showed that the patterning of the facet caused the effective reflectivity to be different for each lateral mode, suggesting that the peak in the measured spectra is associated with the single lateral mode which is most strongly reflected from the modified facet.

*Correspondence: Email g.r.nash@exeter.ac.uk

Efficient, room temperature, mid-infrared 3-5μm semiconductor lasers are required for applications such as gas sensing, free space optical communication, healthcare and missile countermeasures. Compressively strained Type I quantum well laser diodes [1] offer a simpler alternative to quantum cascade lasers as they contain far few layers of different semiconductor material. Recently, room temperature broad area lasers (BALs) were demonstrated emitting 190mW of continuous wave power at a wavelength of 3.1μm [2]. For many applications, both high power and single mode operation are desirable. However, the emission properties of BALs are determined by the complex interplay between the laterally extended waveguide (typically 10s or 100s of microns wide) and the nonlinear, local interaction of the intense light field with the semiconductor active medium, in which gain and refractive index are strongly coupled (leading to filamentation). Overall this leads to a broad optical spectrum with a complex mode structure, which is dependent on the drive current and temperature (together with an output beam that is irregular and multilobed). Methods of suppressing filamentation and controlling lateral modes in laser diodes emitting at shorter wavelengths have therefore been the subject of much investigation for over thirty years. These methods include injection locking techniques [3], the use of external cavities [4], and the spatial control of the reflectivity of the emitting facet [5-9]. In this paper we demonstrate single lateral mode operation of an edge-emitting mid-infrared laser diode using sub-wavelength spatial modulation of the reflectivity of the emitting facet. This approach could ultimately be combined with recent work exploiting plasmonic phenomena in similar sub-wavelength metallic structures [10]. Much of this work has so far concentrated on beam shaping rather than laser modal control. However, linking these two areas of research could open up exciting new avenues for exploring and controlling the processes that govern both the generation of light within, and the emission from, the cavity.

Laser diodes were grown by molecular beam epitaxy onto on-axis semi-insulating (SI) (001) GaAs substrates. The diode structure, shown schematically in Figure 1(a), consists of $Al_{0.3}In_{0.7}Sb$ cladding regions, $Al_{0.11}Ga_{0.17}In_{0.72}Sb$ barriers and two strained $Ga_{0.19}In_{0.81}Sb$ quantum well (QW) active regions [11]. A number of devices, each containing five ridge waveguide laser diodes with sloping sidewalls and a width of ~25µm at the active region, were fabricated using contact photolithography and wet chemical etching. Sputtered Ti/Au metallic contacts were deposited in a 8µm wide dielectric window on top of the ridge, to provide a p-type contact, and on top of the etched region adjacent to the ridge to provide a n-type contact. The devices were cleaved to cavity lengths of 2mm and hand soldered, using indium, substrate-side-down onto coated copper blocks. Further details of the growth, fabrication, and associated characterization are given in ref. 11. Two devices were then characterized before the emitting facet was modified. Sputter-coating was used to deposit 710nm of silicon nitride (for electrical isolation) and 63nm of gold only on the front facets (with the rest of the device masked using photoresist). Focused ion beam (FIB) milling was used to create an array of rectangular 1.7µm wide, 2µm high and 1µm deep apertures in the gold/dielectric, as shown in the image in Figure 1b, on two lasers on each device. The holes were aligned vertically so that the center of the holes coincided with the quantum well active region, and the horizontal separation of the holes was 1.7µm. To enable direct comparison before and after modification of the facets, all measurements were undertaken at 100K with devices mounted on the cold finger of a continuous flow liquid helium cryostat. Three of the lasers were successfully tested after modification of the facets. Lasers were driven with a 1 kHz square wave with a 0.1% duty cycle (pulse length ~ 1 µs) and with a peak current of 2A (high above the typical threshold currents of these lasers).

The measured spectrum from one laser diode before modification of the emitting facet is shown in Figure 2a. Spectra were acquired using a Jobin-Yvon iHR550 grating spectrometer, together with a cooled InSb detector. Before modification of the facet, the measured spectrum consists of a number of significant peaks, with a separation of 2-20nm, consistent with the presence of a number of spatial modes (longitudinal and lateral) and the interplay between them. Figure 2b shows the measured spectrum of the same laser after modification of the emitting facet. This spectrum is radically different from that of the unmodified laser and consists primarily of one large peak, at wavelength of approximately 3362nm, which contains approximately 60% of the total emitted power (the patterning of the facet caused a significant reduction in the number of peaks in the spectra of the three lasers measured, although the effect was most pronounced in this particular laser).

In order to understand these results a Finite Difference Time Domain (FDTD) model was developed to calculate the reflection of different lateral laser modes from the patterned facet. A commercial code from Lumerical was used which can accurately describe wavelength dependence of the real and imaginary parts of the refractive index of the metal layer. This code also allows modal excitation to be performed whereby the allowed waveguide modes present in the laser can be studied individually. The simulation was performed in 2D and used the effective index method to model the vertical confinement in the structure. The lateral guiding is produced by gain guiding and thermal effects [12, 13]. The width of the guiding region is thus related to the current and temperature profile within the laser and was estimated to be equal to 10μm for this device. Thus a 2D waveguide was defined with a central guiding region which was 10μm wide with a refractive of n=4, this was surrounded by cladding regions of n=3.93 (These values of refractive index are typical for InSb based systems in the 3-5μm region [14]). The polarization of the laser emission was using a wire grid polarizer and found that the majority (> 90%) of the electric field is in the direction parallel to the growth planes (the x-direction in Fig. 1a), so only Transverse Electric (TE) modes with this polarization were considered (the polarization was unchanged after modification of the facet).

Mode solving was then performed within Lumerical and 5 TE modes were obtained at $\lambda$=3.4µm. The silicon nitride layer (with n=2) and the metal grating structure were then added at the end of the laser, and each supported lateral mode was made incident on the grating and the total power reflected back into the laser cavity was calculated, as shown schematically in the inset of Figure 3. The reflection monitor was placed a distance of 5µm from the gold-air interface. The calculated reflectance (normalized to the source power) for each lateral TE mode at $\lambda$=3.4µm is plotted in Figure 3. The effective reflectivity of the metallic grating is highest for the fourth TE mode (84% compared to 74% for the fifth mode) suggesting that the patterning of the emitting facet has caused this mode to become dominant. This is consistent with the measured emission spectrum, where the peak after modification of the facet occurs towards the shorter wavelength end of the range of wavelengths observed before modification, as might be expected if this dominant spectral mode is associated with a higher order lateral mode. The power emitted from the front facet was measured using a calibrated cooled InSb detector mounted on an x-y stage and, correcting for the detector response and geometry, was ~32mW and ~9mW before and after the facet modification respectively. The decrease in emitted power is consistent with an increase in the effective reflectivity of the emitting facet reflectivity, which was approximately 33% before modification (note that the back facet was unmodified). Further work is underway both to simulate the far field intensity profiles and also to investigate the use of different aperture patterns to allow the selection of particular spatial modes.

In conclusion we have investigated the characteristics of mid-infrared laser diodes before and after the patterning of sub-wavelength metallic apertures on the emitting facet. Before modification of the facet the emitted spectrum consisted of a large number of peaks, associated with the different spatial modes supported, whereas afterwards the spectrum was dominated by a single peak. Simulations carried out using a Finite Difference Time Domain (FDTD) model showed that the patterning of the facet caused the effective facet reflectivity to be different for each of the lateral modes supported, suggesting that the peak in the measured spectra is associated with the single lateral mode which is reflected most strongly from the modified facet. There is significant potential to combine this approach with recent work which has used plasmomic phenomena to control the properties of the emitted beam.

## ACKNOWLEDGMENTS

This work was supported by the UK Centre for Defence Enterprise and EPSRC, and the authors wish to thank Professor David Titterton, Professor Mike Jenkins and Dr Ewan Finlayson for useful discussions.

**FIGURE CAPTIONS**

Figure 1.  (a) Schematic cross-section of the laser diodes and (b) image taken in the Focused Ion Beam system showing the array of apertures patterned in the metal. The arrow in (b) indicates the approximate vertical position of the quantum wells.

Figure 2.  Measured emission spectra (a) before and (b) after modification of the emitting facet.

Figure 3.  Reflectance of each lateral TE mode.  The inset shows a schematic representation of the model used to calculate the reflectivity (RM: reflection monitor, MS: mode source, GG: gold grating).

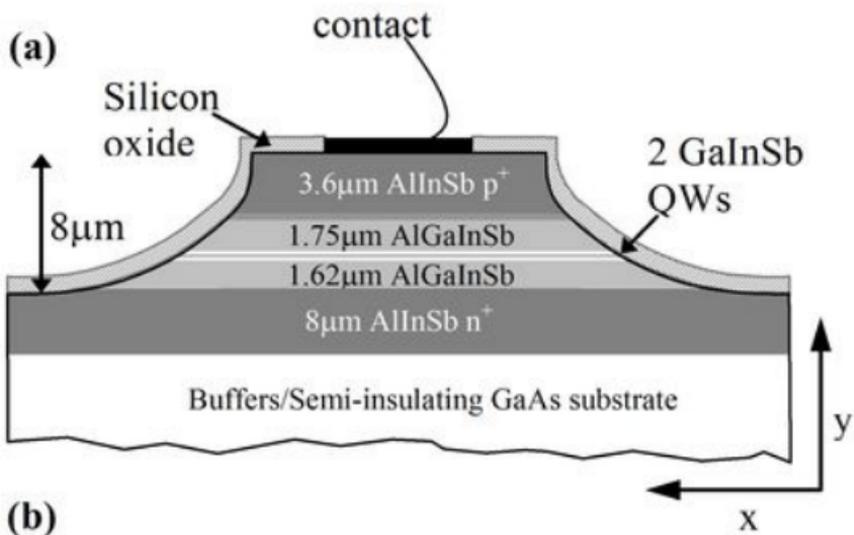

**(a)**

contact

Silicon oxide

2 GaInSb QWs

8μm

3.6μm AlInSb p$^+$

1.75μm AlGaInSb

1.62μm AlGaInSb

8μm AlInSb n$^+$

Buffers/Semi-insulating GaAs substrate

y

x

**(b)**

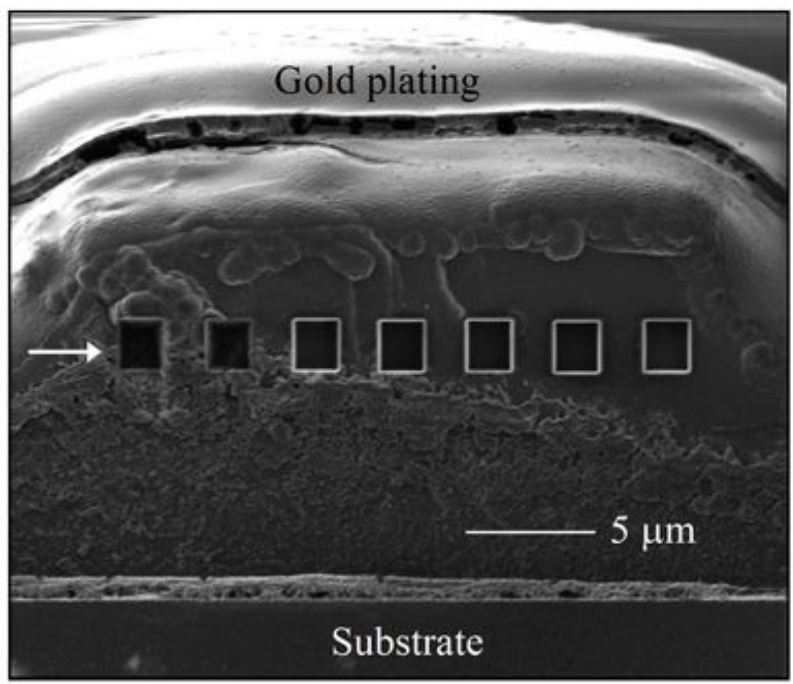

Gold plating

5 μm

Substrate

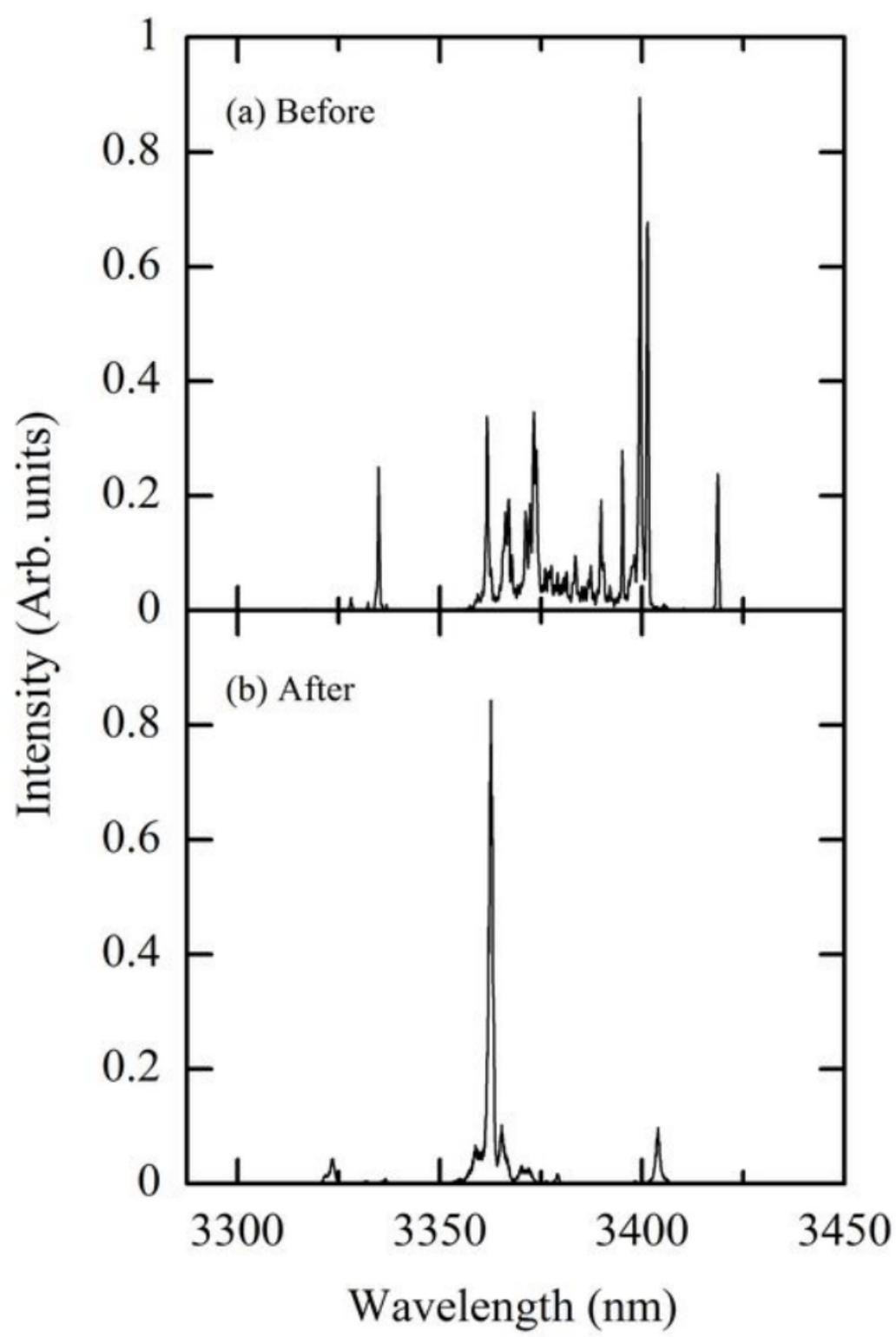

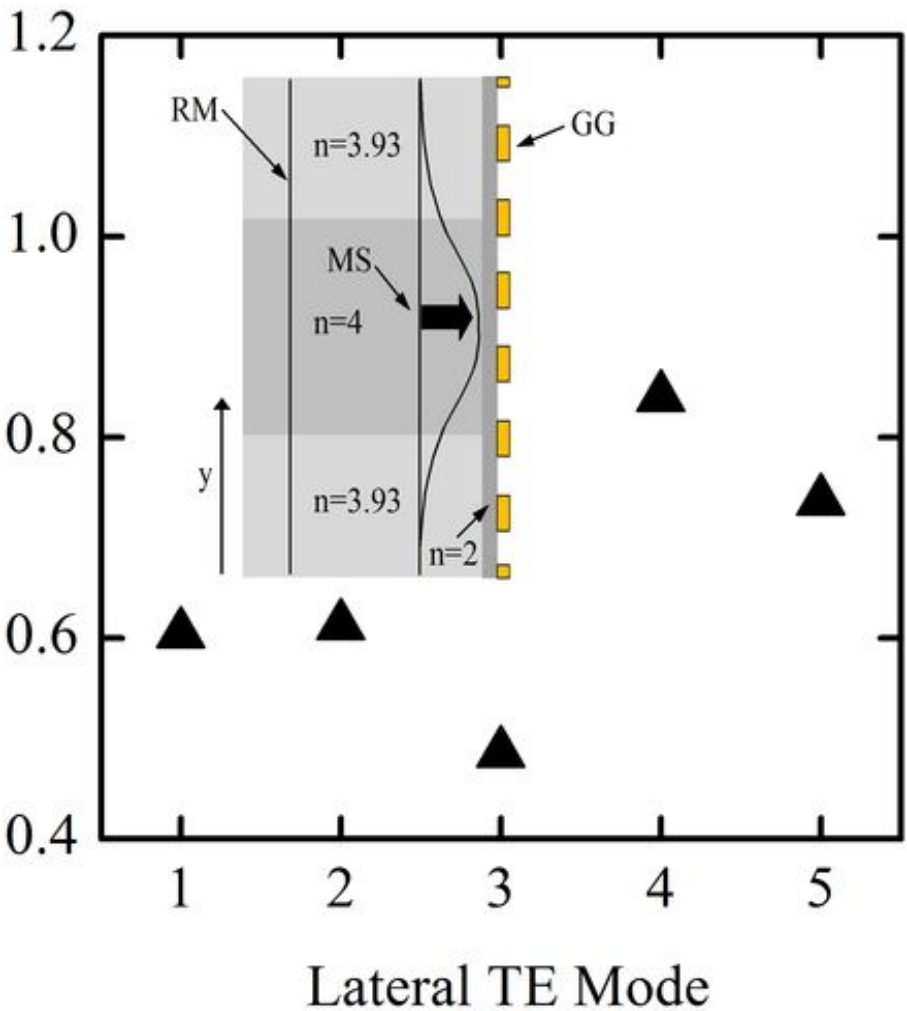